\documentclass[a4paper,11pt]{article}
\usepackage{pos}

\title{Testing the influence of anisotropic CR transport and the Galactic magnetic field structure on the all-sky gamma-ray emission}
\ShortTitle{Gamma-rays from anisotropic CR transport}

\author*[a,b]{Julien Dörner}
\author[a,b]{Jonas Hellrung}
\author[a,b,c]{Julia Becker Tjus}
\author[a,b]{Horst Fichtner}

\affiliation[a]{Theoretische Physik IV, Fakult\"at f\"ur Physik \& Astronomie, Ruhr-Universit\"at Bochum, 44780 Bochum, Germany}
\affiliation[b]{Ruhr Astroparticle and Plasma Physics Center (RAPP Center), Germany}
\affiliation[c]{Department of Space, Earth and Environment, Chalmers University of Technology, 412 96 Gothenburg, Sweden}

\emailAdd{jdo@tp4.rub.de}

\abstract{
The spatial diffusion of energetic particles in a magnetic field composed of a large-scale background and a small-scale turbulent component should be expected to be anisotropic. While such anisotropic diffusion has been known for quite a while in first-principle plasma physics and while it is required for an understanding of the transport of cosmic rays in the heliosphere or close to supernova remnants, only in recent years it has also become of particular interest for the modeling of Galactic cosmic ray (GCR) transport in the Milky Way in the context of their residence time and their (local) energy spectra. Also, the large-scale spatial distribution of GCRs is shaped by an anisotropic diffusion in the Galactic magnetic field, which should directly affect both the diffuse gamma-ray and the neutrino emission.

We solve the anisotropic diffusive transport of GCRs in the Milky Way using the publicly available transport code CRPropa.  The anisotropy of the diffusion is characterized by the ratio between the diffusion coefficient perpendicular and parallel to the local magnetic field $\epsilon = D_\perp / D_\parallel$, where we test different values reaching from nearly parallel transport ($\epsilon = 10^{-3}$) to more isotropic diffusion ($\epsilon = 10^{-1}$). 

From the three dimensional distribution of GCRs in the Milky Way we calculate the all-sky gamma-ray emission, using the line-of-sight integration framework HERMES. Finally, we demonstrate the impact of the anisotropy in the diffusion on the spatial distribution of the gamma-ray flux and its spectral energy distribution. It shows strong influences by the anisotropy of the diffusion and the magnetic field geometry.
}

\ConferenceLogo{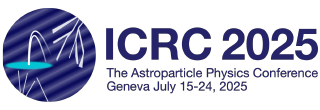}

\FullConference{39th International Cosmic Ray Conference (ICRC2025)\\
 15–24 July 2025\\
Geneva, Switzerland\\}
\begin{document}
\maketitle

\section{Introduction}

Galactic Cosmic Rays (GCRs) are accelerated at local sources in the Galaxy, travel through interstellar space, and interact with the ambient matter and photon fields, leading to non-thermal emission (see \cite{Tjus2020} for a review). The interactions with the ambient matter are the dominant source for the diffuse all-sky gamma-ray emission.  
For a precise description of these gamma rays, a detailed knowledge of the underlying GCR transport processes and the interaction is needed. 

It has been shown that the modeling of the interaction itself leads to systematic uncertainty of a factor $\sim 2$ in the total brightness while the spectral slope is nearly unaffected \cite{Doerner25}. In the presence of a large-scale background magnetic field in addition to the turbulent component, the transport of GCRs is expected to become anisotropic with respect to the background field direction \cite{Reichherzer21}. This aspect has become of particular interest in the past decade for modeling Galactic transport and its non-thermal emission \cite{Effenberger12, Giacinti13, Pakmor16 ,Cerri17}.

The first idea to constrain the anisotropy of the diffusion tensor with gamma ray observations in a compact source region is presented in \cite{Doerner24}. Here, the Central Molecular Zone (CMZ) of the Milky Way has been modeled and compared to the gamma-ray observation by the H.E.S.S. telescope. The anisotropy of the diffusion tensor was introduced by the ratio between the diffusion coefficient perpendicular and parallel to the background magnetic field $\epsilon = D_\perp / D_\parallel$. In the CMZ, the best agreement to the observational data is achieved for nearly isotropic diffusion ($\epsilon \approx 1$). In this work, we extend this work to cover the full Milky Way and predict differences in the all-sky gamma ray emission from the anisotropy in the diffusion tensor and the geometry of the Galactic Magnetic Field (GMF).

\section{Galactic Magnetic Field}
Several attempts have been made to develop a model for the GMF. One of the first realistic models was developed by Jansson \& Farrar \cite{JF12_orig}, which has been widely used in cosmic-ray physics. The poloidal X-shape and the spiral disc component of this field model contain regions in which the divergence constraint is violated. For this purpose, the authors of \cite{JF12_sol} proposed a smoothing to address this problem. In the following, we refer to the solenoidal improved version as JF12. 

\begin{figure}[t]
    \centering
    \includegraphics[width=.495\linewidth]{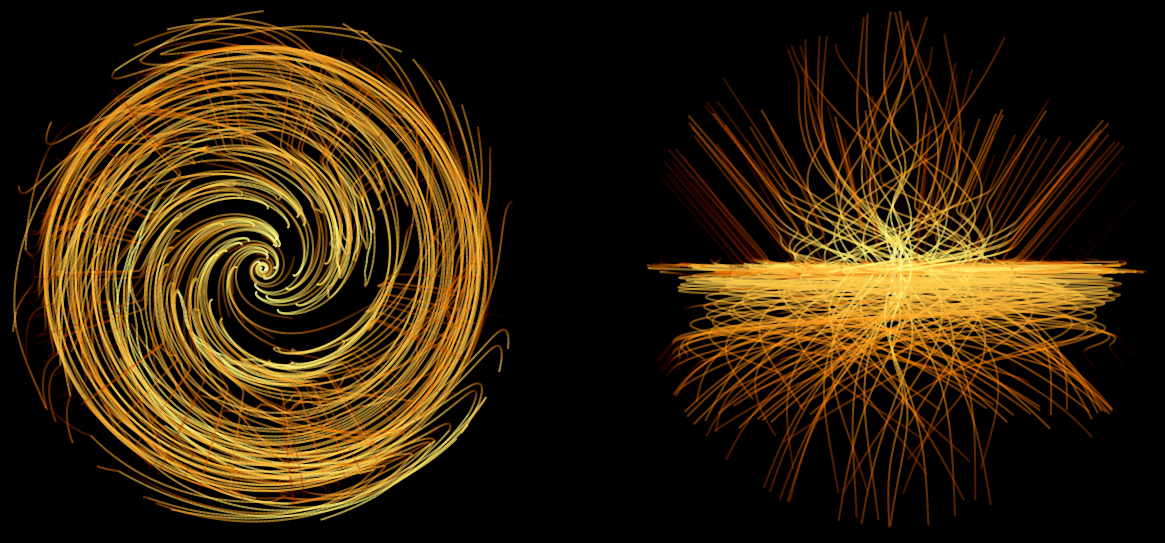}
    \includegraphics[width=.495\linewidth]{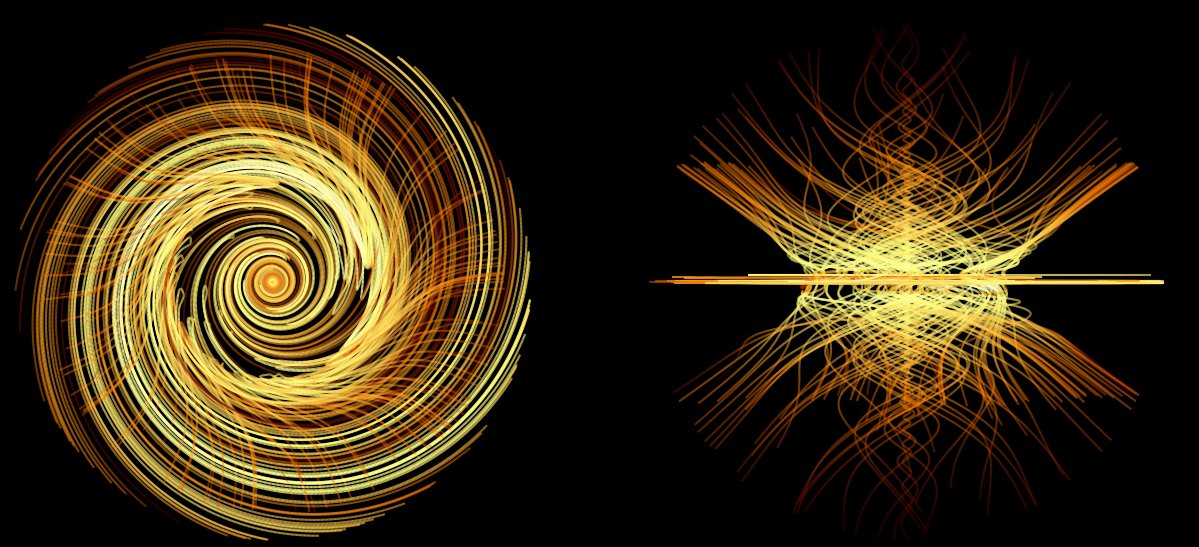}
    \caption{Fieldlines in the solenoidal version of the JF12 field (left panel) \cite{JF12_sol} and the baseline model of UF23 (right pannel) \cite{UF23} in a face-on and edge-on view.}
    \label{fig:fieldlines}
\end{figure}

Any GMF model is based on several assumptions, like the geometry, the distribution of thermal and non-thermal electrons. Unger \& Farrar \cite{UF23} derive a set of different GMF models relying on different assumptions, leading to a similar agreement with observation data. In Fig.~\ref{fig:fieldlines} a set of field lines in the edge-on and face-on view for the JF12 and the baseline model of UF23 are shown. These GMF models show significant differences in the shape, which are expected to lead to differences in the CR distribution considering anisotropic diffusion. In the following, we test the anisotropy of the diffusion tensor in the UF23 baseline model. In the future, we will extend this work to include the other GMF models as well.

\section{CR transport in the Milky Way}
From the observation of the GCR proton spectrum, several breaks in the power-law slope are known \cite{Tjus2020}. Observations of the B/C ratio indicate that the break at around hundred GeV is caused by the diffusion of GCRs, which can be explained by the \textit{streaming instability}, where GCRs scatter on waves excited by themselves \cite{Blasi12}. In a one-dimensional approximation, the coupled wave-spectrum and GCR distribution can be derived, and the effective diffusion coefficient can be extracted, where its energy dependence is described by a broken power-law 
\begin{equation}
    D_\parallel(E) = D_0 \, \left(\frac{E}{E_{br}}\right)^{\gamma_1} \, \left( 1 + \left( \frac{E}{E_{br}}\right)^{\gamma_2 - \gamma_1} \right)^{-1} \quad .
\end{equation}
In \cite{Doerner_PHD} this non-linear model was fitted to available GCR data, resulting in a break at $E_{br} = (64.38 \pm 0.11)$ GeV, assuming a \textit{Kolmogorov}-like energy dependence above the break ($\gamma_1 = 0.335$ and $\gamma_2 = -0.321$). In the following, we apply this fit to all simulations. We note that this approximation is only valid for GCR energies $\gtrsim 10$ GeV and restrict our model to this energy range. 

The one-dimensional model of the \textit{streaming instability} gives access to the diffusion coefficient parallel to the magnetic field line. To include the effect of diffusion perpendicular to the field line, we define the anisotropy of the diffusion $\epsilon$ by the ratio between the perpendicular and the parallel diffusion coefficient $\epsilon = D_\perp / D_\parallel$. The expected value for this anisotropy depends, among others, on the ratio between the background magnetic field $B$ and the turbulent component $b$, which is not known for the Milky Way. Therefore, we test different scenarios $\epsilon = 10^{-1}, 10^{-2}, 10^{-3}$, to investigate the impact of the anisotropic diffusion on the three-dimensional CR distribution and subsequently on the gamma ray emission. 

\subsection{Modeling the CR distribution with CRPropa}
To model the 3+1-dimensional distribution of GCR protons in space and energy, we use the publicly available simulation framework CRPropa \cite{CRPropa3.2}. The method of Stochastic Differential Equations (SDEs) \cite{CRPropa3.1} is applied to model the diffusive behavior of the GCRs. Following the idea of \cite{CRPropa3.1}, we calculate the steady-state solution by summing up snapshots of a burst injection. As the escape time of GCRs varies with energy, the simulation is split into three different sets, each with a different time resolution for the snapshots. The details about the energy range and the temporal resolution are given in Tab.~\ref{tab:sim_set}. The halo height is fixed to $H = 4$ kpc and the extent of the disk to $R = 20$ kpc. The spatial source distribution follows the pulsar distribution \cite{BlasiAmato12}.

\begin{table}[tb]
    \centering
    \begin{tabular}{c|cccc}
         Parameter & $E_\mathrm{min}$ [GeV] & $E_\mathrm{max}$ [GeV] & time step $\Delta T$ [kpc$/c$] & maximal time $T_\mathrm{max}$ [Gpc$/c$]  \\ \hline
         low & $10^{-1}$ & $10^2$ & $2000$ & $50$  \\
         medium & $10^2$ & $10^5$ & 10 & 1\\
         high & $10^5$ & $10^7$ & 1 & 1
    \end{tabular}
    \caption{Time resolution and energy ranges for the different simulation regimes.}
    \label{tab:sim_set}
\end{table}

\begin{figure}
    \centering
    \includegraphics[width=\linewidth]{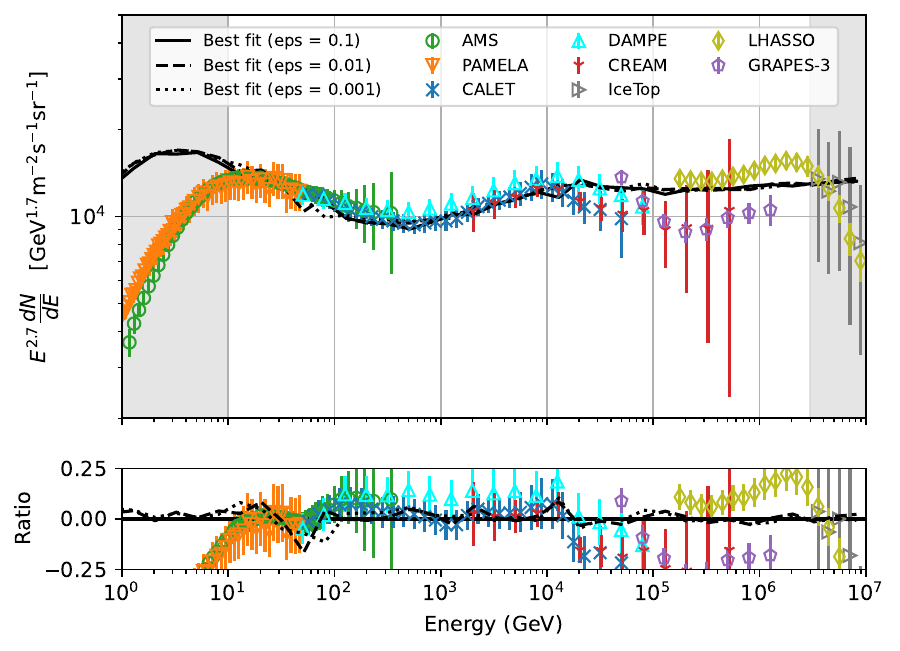}
    \caption{Cosmic Ray proton spectrum arriving at Earth for the optimized injection parameters. The black lines show the different anisotropies in the diffusion. The colored data points show the observations from \cite{AMS, PAMELA, CALET, DAMPE, CREAM, IceTop, GRAPES, LHASSO}. The gray regions are excluded from the fit to avoid contamination from the \textit{knee} at high energies and due to the energy range of the diffusion approximation applied here.}
    \label{fig:opt_CR}
\end{figure}

The simulated energy distribution is flat in $\log(E)$ to achieve equal statistical uncertainties in all logarithmic energy bins. After the simulation, the results can be reweighted to any energy dependence of the source spectrum. From the full simulation set, all CRs within $100$ pc around the position of the Earth are selected to calculate the local interstellar spectrum as it would arrive at Earth. The parameters of a smooth broken power-law in momentum 
\begin{equation}
    Q(p) = Q_o \, \left(\frac{p}{p_{br}}\right)^{\alpha_1} \left[ 1 + \left(\frac{p}{p_{br}}\right)^{w^{-1}}\right]^{(\alpha_2 - \alpha_1) w}
\end{equation}
are fitted for the best agreement to the observed GCR data from AMS \cite{AMS}, PAMELA \cite{PAMELA}, CALET \cite{CALET}, \cite{DAMPE}, ISS-CREAM \cite{CREAM}, IceCube/IceTop \cite{IceTop}, Grapes-3 \cite{GRAPES}, and LHASSO \cite{LHASSO}. The resulting GCR spectra for all simulated anisotropies in the diffusion are shown in Fig.~\ref{fig:opt_CR}.
All simulated anisotropies show a good agreement to the local GCR spectrum and reproduce the break in the energy spectrum at $\sim 10$ TeV as reported by \cite{CALET, DAMPE}. The fit of the local GCR data does not constrain the anisotropy of the diffusion. The best-fit source parameters are afterwards applied to the full dataset to calculate the GCR distribution in the Milky Way. 

\section{Gamma ray emission}
The gamma ray emission from the hadronic interactions of the GCRs within the Milky Way is calculated with the HERMES package \cite{HERMES}. It applies the line-of-sight (LOS) integration
\begin{equation}
    I_\gamma(E_\gamma, l, b) = \int\limits_{0}^{\infty} \mathrm{d}s \, n_H(\vec{r}) \int \mathrm{d}E_p \, \Phi_p(E_p, \vec{r}) \frac{\mathrm{d}\sigma_{pp}}{\mathrm{d}E_\gamma}(E_p, E_\gamma)
\end{equation}
for a given direction $(l,b)$ on the sky. The spatially dependent GCR proton flux $\Phi_p(E_p, \vec{r})$ is taken from the CRPropa simulation discussed before. As gas distribution $n_H(\vec{r})$ the ring models provided in the HERMES package for the atomic $H_I$ and molecular $H_2$ hydrogen are used. 
The differential production cross section $\mathrm{d}\sigma_{pp}/\mathrm{d}E_\gamma$ from AAfrag \cite{AAfrag} is applied.

\begin{figure}
    \centering
    \includegraphics[width=\linewidth]{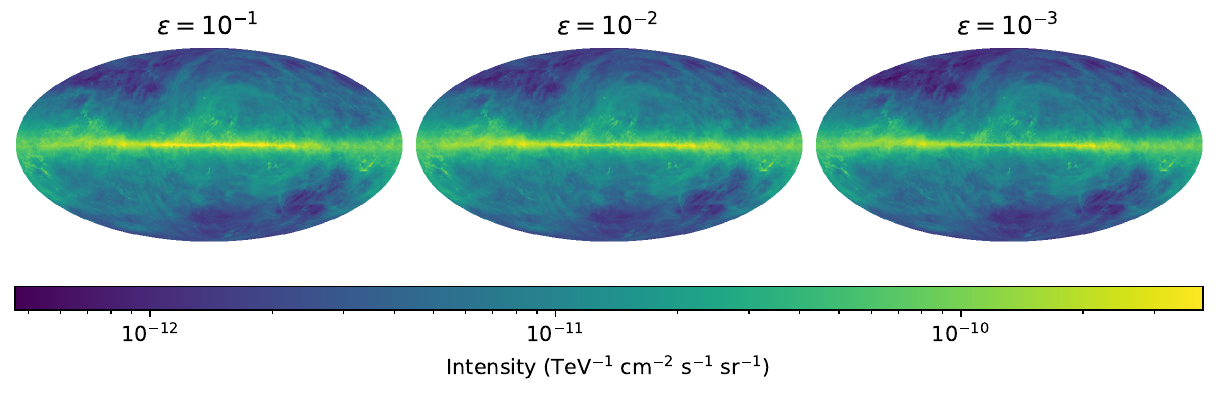}
    \caption{All-sky gamma ray emission for different anisotropies of the CR transport}
    \label{fig:maps}
\end{figure}

\begin{figure}
    \centering
    \includegraphics[width=\linewidth]{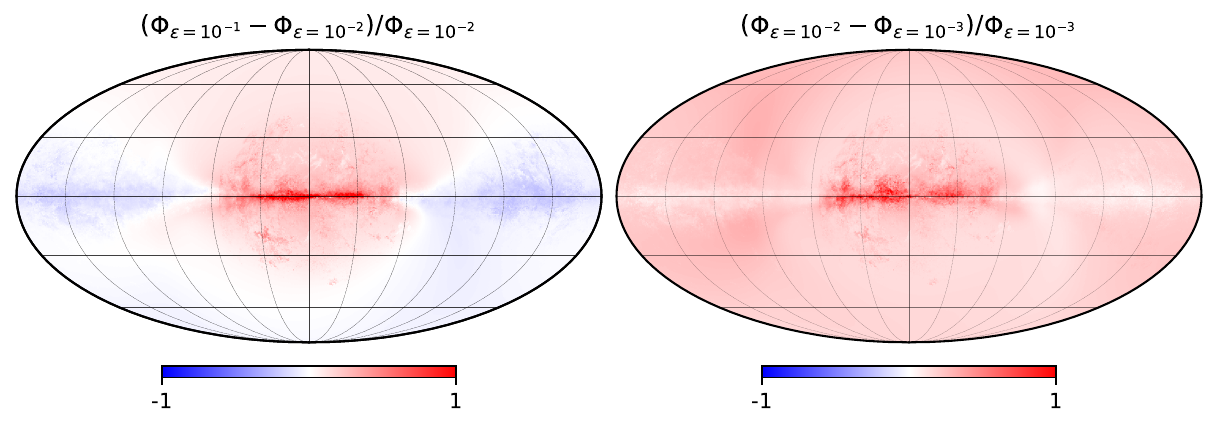}
    \caption{Ratio between the gamma ray emission shown in Fig.~\ref{fig:maps}. Here, the maps are compared to the next stronger anisotropy. Red colors indicate more gamma-ray production in the isotropic diffusion, while blue colors show more gamma rays from the anisotropic simulation.}
    \label{fig:maps_ratio}
\end{figure}

In Fig.~\ref{fig:maps}, the resulting all-sky flux at $E_\gamma = 1$ TeV is shown for the different anisotropies of the CR diffusion. The general structure of the emission stays the same for all anisotropies as the gas distribution shapes it. To illustrate the differences between these maps, the relative difference is shown in Fig.~\ref{fig:maps_ratio}. 

In the more isotropic case ($\epsilon = 0.1$ vs $\epsilon = 0.01$, left panel of Fig.~\ref{fig:maps_ratio}), one can see that the inner Galaxy the predicted flux is higher in the more isotropic diffusion scenario (red color) while it is lower in the outer Galaxy (blue color). The geometry of the magnetic field determines this behavior (compare Fig.~\ref{fig:fieldlines}). In the outer Galaxy, the field lines are mainly oriented within the Galactic disc. Therefore, a strongly anisotropic diffusion will increase the confinement of CRs, and the gamma-ray production is increased. In the inner Galaxy, more field lines are pointing away from the disc, leading to faster CR escape in the parallel-dominated diffusion. Therefore, the more anisotropic diffusion ($\epsilon = 10^{-2}$) leads to a deficit in produced gamma rays.

In the case of extremely anisotropic transport ($\epsilon = 10^{-3}$, right panel of Fig.~\ref{fig:maps_ratio}), the overall gamma-ray emission at 1 TeV is suppressed. The GCRs are strongly confined to the magnetic field lines and as the GCR sources are only located within the spiral arms the particles can not propagate into the region in between. Therefore, the production of gamma-rays, which is integrated over the full line-of-sight, is reduced. Besides the overall offset in the ratio, the spatial structure stays the same when comparing the ratios in Fig.~\ref{fig:maps_ratio}, as this is shaped by the magnetic field geometry. 

\begin{figure}
    \centering
    \includegraphics[width=\linewidth]{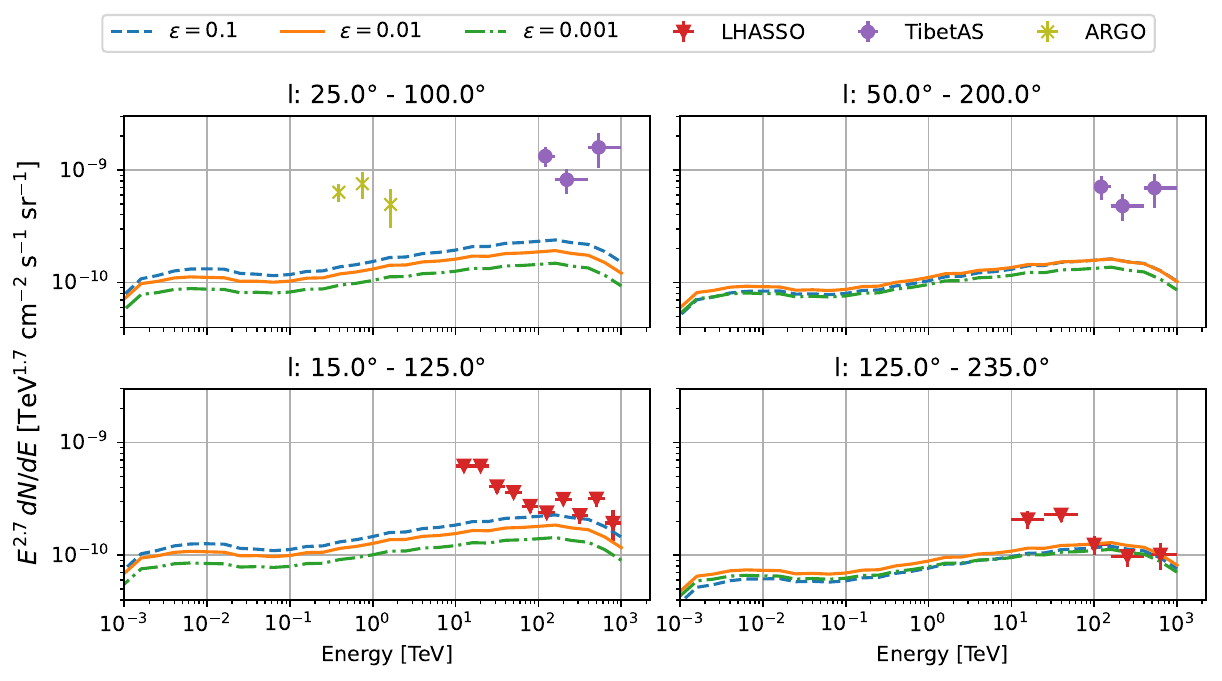}
    \caption{Spectral energy distribution of the gamma-ray emission in different sky regions from the anisotropic CR diffusion models compared to the data from \cite{LHASSO_gamma, Tibet, ARGO}.}
    \label{fig:SED_gamma}
\end{figure}

To get a more quantitative comparison with the observed diffuse gamma-ray emission, we select four different regions within the Galactic disc ($|b| \leq 5^\circ$), two in the inner and two in the outer. The regions are chosen to match the sky coverage of the observations by TibetAS$\gamma$ \cite{Tibet} (inner: $25^\circ \leq l \leq 100^\circ$, outer: $50^\circ\leq l \leq 200^\circ$), ARGO \cite{ARGO} and LHASSO \cite{LHASSO_gamma} (inner: $15^\circ \leq l \leq 125^\circ$, outer: $125^\circ \leq l \leq 235^\circ$). While the observations mask the point sources in their observational window, no additional blinding is applied for the simulations. The full SED in these regions is shown in Fig.~\ref{fig:SED_gamma}. While the spectra for the outer Galaxy (right column) do not show large deviations between the different anisotropies, the differences in the inner Galaxy are much stronger. 

All predictions are slightly below the measured fluxes, which can be expected as the model presented here, contains only protons. Additional contribution from heavier elements is expected to be at the order of a factor at most $\sim 2$ of the proton prediction \cite{EspinosaCastro25}, considering the LHASSO proton and all-particle flux. At these energies contribution from leptonic processes, which have not been included so far, can be neglected. Only at the highest energies $E\gtrsim 100$ TeV the model expectation becomes comparable with the observation. Here, we have to note that the well-known CR \textit{knee} at $\sim 3$ PeV is not taken into account for the underlying CR distribution (see Fig.~\ref{fig:opt_CR}). If we consider an average energy transfer of 10\% into gamma-rays in hadronic interaction, one would expect a spectral break at about $E_\gamma \approx 300$ TeV, which could allow for the contribution by heavier elements also in this energy range. 

\section{Summary and discussion}
The diffusion of cosmic rays in a magnetic field composed of a large-scale background magnetic field and a turbulent component is expected to be anisotropic. This anisotropy can be characterized by the ratio between the diffusion coefficient perpendicular and parallel to the background field direction $\epsilon = D_\perp / D_\parallel$. We solved the anisotropic CR transport in the Milky Way using the baseline model from \cite{UF23} as the large-scale GMF and testing different values for $\epsilon$. The CR source spectrum is
fitted to the observed CR proton flux at Earth, including a break in the injection spectrum at $\sim$ 10 TeV. All anisotropy levels of the diffusion can reproduce the observed CR data. 

From the results of the 3+1-dimensional solution of the CR transport we calculated the all-sky gamma ray emission due to hadronic interaction of the CR protons with the ambient gas. 
While the over-all shape of the emission is dominated by the gas distribution, small changes depending on the anisotropy of the diffusion tensor can be observed. Specifically, the inner and the outer Galactic disc show different gamma-ray intensities depending on the anisotropy as the magnetic field configuration changes from pointing out of the disc in the inner plane to in-plane field lines for the outer disc. 

The observed gamma-ray spectra are in agreement with the prediction from the anisotropic simulations, and additional contribution from heavier nuclei is expected. In the future, one has to take the CR \textit{knee} into account to include the spectral change in the observed gamma-ray spectrum at several hundred TeV. A direct comparison of the predicted gamma ray maps with all-sky observations from e.g., Fermi-LAT at the lower energies and LHASSO at the highest energies will help to constrain the anisotropy of the diffusion even further. 

Besides non-thermal gamma ray emission one would also expect a neutrino signal to arise from these hadronic interactions. Observations of the IceCube observatory show emission from the Galactic Plane\cite{IceCubePlane23}. The calculations presented here, can be extended to include the neutrino production and a comparison to the IceCube observations may help to give better constraints on the anisotropic CR diffusion.

\section*{Acknowledgments}
The authors acknowledge funding from the German Science Foundation DFG, via the Collaborative Research Center SFB1491 - Cosmic Interacting Matters - From Source to Signal.

\setlength{\bibsep}{0pt plus 0.3ex}

\end{document}